\documentclass[a4paper]{jpconf}
\usepackage{graphicx}
\usepackage{amsmath,amssymb}
\usepackage{cite}
\usepackage[dvips]{color}

\begin{document}
\title{Gate-voltage dependence of Kondo effect in a triangular quantum dot}

\author{T Numata$^1$, Y Nisikawa$^1$, A Oguri$^1$, and A C Hewson$^2$}

\address{$^{1}$ Department of Material Science, Osaka City University, 
Osaka 558-8585, Japan}

\address{$^{2}$ Department of Mathematics, Imperial College, 
 180 Queen's Gate, London SW7 2BZ, UK}

\ead{oguri@sci.osaka-cu.ac.jp}

\begin{abstract}
We study the conductance through a triangular triple quantum dot, 
which are connected to two noninteracting leads, 
 using the numerical renormalization group (NRG).
It is found that the system shows a variety of Kondo effects 
depending on the filling of the triangle. 
The SU(4) Kondo effect occurs at half-filling, 
and a sharp conductance dip due to a phase lapse 
appears in the gate-voltage dependence.
Furthermore, when four electrons occupy the three sites on average, 
a local $S=1$ moment, which is caused by the Nagaoka mechanism, 
is induced along the triangle. 
The temperature dependence of the entropy 
and spin susceptibility of the triangle shows that  
this moment is screened by the conduction electrons via 
two separate stages at different temperatures. 
The two-terminal and four-terminal conductances 
show a clear difference at the gate voltages,  
where the SU(4) or the $S=1$ Kondo effects occurring.

%
% The half of the $S=1$ moment is screened first by one 
% of the channel degrees, 
% and then at very low temperature the remaining half is screened 
% to form a fully screened Kondo singlet. 
%

\end{abstract}

\section{Introduction}

% The Kondo effect in quantum dots 
% is an active field of current research,
% and recently 
%

The Kondo effect in quantum dots 
has been studied in various kinds of systems in recent years, 
and the triple quantum dots with a triangular configuration 
is one of the interesting systems \cite{Vidan-Stopa,TKA}. 
Specifically, the closed path along the triangle 
causes a high-spin ground state with the Nagaoka mechanism \cite{ONTN}, 
and also an SU(4) Kondo effect, depending on the gate voltage. 
In this report we study the triangular triple dot connected to 
two noninteracting leads, 
and present the results of the scattering phase shifts 
obtained with the numerical renormalization group (NRG). 
%  From the phase shifts, the series and parallel conductances 
% for the current flowing along the horizontal direction 
% in the configurations shown in Fig.\ \ref{fig:system} can be  deduced. 
We also calculate the temperature dependence 
of the entropy and spin susceptibility \cite{KWW}, 
which show clearly the two-stage screening process 
of the high-spin Nagaoka state and 
also the feature of the Kondo screening in the SU(4) 
case at half-filling.
% 

% \cite{NO,ONH,OH}. 

% $i$) how the Nagaoka ferromagnetism 
% that could manifest in the isolated triangle 
% for a particular charge filling is screened 
% by the conduction electrons, and 
% $ii$) how it affects the low-temperature transport bellow 
% the Kondo energy scale $T_K$, 
%  we present the results using 
% the NRG approach 

% which is applicable to low-temperatures $T \lesssim T_K$. 
 
%

\section{Model \& Formulation}

\begin{figure}[t]
\begin{center}
\setlength{\unitlength}{0.7mm}

%\rule{1cm}{0cm}
\begin{minipage}[t]{7cm}

\begin{picture}(110,20)(0,0)
\hspace{-0.8cm}
\thicklines

\put(8.0,7){\makebox(0,0)[bl]{\large (a)}}
%\put(8.5,7){\makebox(0,0)[bl]{\large (a)}}

\put(19,6){\line(1,0){20}}
\put(19,14){\line(1,0){20}}
\put(39,6){\line(0,1){8}}

\put(71,6){\line(1,0){20}}
\put(71,14){\line(1,0){20}}
\put(71,6){\line(0,1){8}}

\multiput(40.0,10)(2,0){4}{\line(1,0){1}}
\multiput(62.9,10)(2,0){4}{\line(1,0){1}}

\put(55,18.6){\circle*{3.3}} 
\put(50,10){\circle*{3.3}} 
\put(60,10){\circle*{3.3}} 

\put(51,10){\line(1,0){8}}
\put(49.5,10){\line(3,5){6}}
\put(60.5,10){\line(-3,5){6}}

\put(42.5,5){\makebox(0,0)[bl]{\large $v$}}
\put(65,5){\makebox(0,0)[bl]{\large $v$}}

%\put(42.5,6){\makebox(0,0)[bl]{\large $v$}}
%\put(65,6){\makebox(0,0)[bl]{\large $v$}}

\end{picture}
\end{minipage}
%
%% \end{center}
%% \caption{Schematic picture of a series connection }
%% \label{fig:series}
%% \end{figure}
%% 
%-------------------------------------------------------------------
%% 
%% 
%% \begin{figure}[tb]
%% \begin{center}
% \setlength{\unitlength}{0.75mm}
\setlength{\unitlength}{0.7mm}
%
%\rule{1cm}{0cm}
\begin{minipage}[t]{7cm}

\begin{picture}(110,30)(-4,0)
\hspace{-0.8cm}
\thicklines

\put(4.5,13){\makebox(0,0)[bl]{\large (b)}}
%\put(4.75,13){\makebox(0,0)[bl]{\large (b)}}

\put(16.5,4.5){\line(1,0){20}}
\put(16.5,12.5){\line(1,0){20}}
\put(36.5,4.5){\line(0,1){8}}

\put(16.5,17.5){\line(1,0){20}}
\put(16.5,25.5){\line(1,0){20}}
\put(36.5,17.5){\line(0,1){8}}

\put(67.5,4.5){\line(1,0){20}}
\put(67.5,12.5){\line(1,0){20}}
\put(67.5,4.5){\line(0,1){8}}

\put(67.5,17.5){\line(1,0){20}}
\put(67.5,25.5){\line(1,0){20}}
\put(67.5,17.5){\line(0,1){8}}

\multiput(38.0,21)(2,0){5}{\line(1,0){1}}
\multiput(38.0,9.5)(2,0){5}{\line(1,0){1}}
\multiput(53.2,9.5)(2,0){7}{\line(1,0){1}}
\multiput(53.2,21)(2,0){7}{\line(1,0){1}}

\put(58.6,15){\circle*{3.3}} 
\put(50,10){\circle*{3.3}} 
\put(50,20){\circle*{3.3}} 

\put(50,10){\line(0,1){10}}
\put(51,10){\line(3,2){7}}
\put(51,20){\line(3,-2){7}}

\put(37.5,22.5){\makebox(0,0)[bl]{$v/\sqrt{2}$}}
\put(54,22.5){\makebox(0,0)[bl]{$v/\sqrt{2}$}}

\put(37.5,1){\makebox(0,0)[bl]{$v/\sqrt{2}$}}
\put(54,1){\makebox(0,0)[bl]{$v/\sqrt{2}$}}

% \put(37.5,2){\makebox(0,0)[bl]{$v/\sqrt{2}$}}
% \put(54,2){\makebox(0,0)[bl]{$v/\sqrt{2}$}}

\end{picture}
\end{minipage}
\end{center}
\caption{Triangular triple quantum dot 
in (a) series and (b) parallel configurations}
\label{fig:system}
\end{figure}
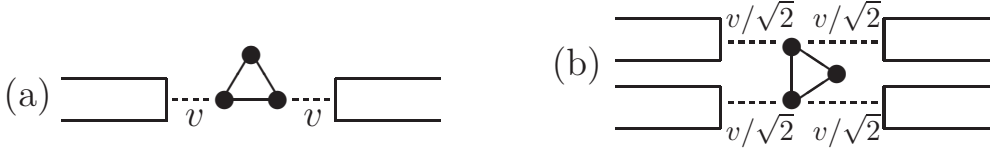

We consider a three-site Hubbard model on a triangle,
which is connected to two non-interacting leads 
on the left ($L$) and right ($R$), 
at $i=1$ and $i=N_D$ ($\equiv 3$), respectively, 
as shown in Fig.\ \ref{fig:system} (a).
The total Hamiltonian has the form   
$\ \mathcal{H} =  \mathcal{H}_{D} +  \mathcal{H}_{\rm mix}  +  
\mathcal{H}_{\rm lead}\,$, with 
\begin{align}
& 
\!\!\!\!\!
\mathcal{H}_{D}  \,= \,    
 -t 
 \sum_{<ij>}^{N_D}\sum_{\sigma}  
 \left(\,
 d^{\dagger}_{i\sigma}d^{\phantom{\dagger}}_{j\sigma}  
+ d^{\dagger}_{j\sigma}d^{\phantom{\dagger}}_{i\sigma}  
\right) 
% \nonumber
% \\
+ \,
\epsilon_d 
\sum_{i=1}^{N_D}\sum_{\sigma} \, 
 d^{\dagger}_{i\sigma}d^{\phantom{\dagger}}_{i\sigma}  
 +  U\sum_{i=1}^{N_D} 
   d^{\dagger}_{i \uparrow}
   d^{\phantom{\dagger}}_{i \uparrow}
   d^{\dagger}_{i \downarrow}
   d^{\phantom{\dagger}}_{i \downarrow} \;, 
\label{eq:HC^U}
\\
&
\!\!\!\!\!
\mathcal{H}_{\rm mix} \,= \,  
v  \sum_{\sigma}   
 \left(  \,
  d^{\dagger}_{1,\sigma} C^{\phantom{\dagger}}_{L \sigma}
% +  
% \psi^{\dagger}_{L \sigma}   d^{\phantom{\dagger}}_{1,\sigma} 
% \,\right)
% \nonumber 
% \\
%  + v 
%  \sum_{\sigma}   
% \left( \,
% \psi^{\dagger}_{R\sigma} d^{\phantom{\dagger}}_{N_D, \sigma} 
   +   
d^{\dagger}_{N_D, \sigma} C^{\phantom{\dagger}}_{R\sigma}
+\, \mathrm{H.c.}
\, \right)   ,
\label{eq:Hmix}
\qquad \quad 
\mathcal{H}_{\rm lead}  \,=  
\sum_{\nu=L,R} 
 \sum_{k\sigma} 
  \epsilon_{k}^{\phantom{0}}\,
         c^{\dagger}_{k \nu \sigma} 
         c^{\phantom{\dagger}}_{k \nu \sigma}
\,.
% \rule{2.6cm}{0cm} 
% \label{eq:H_lead}
\end{align}
%%%%%%%%%%%%%%%%%%%%%%%%%%%%%%%%%%%%%%%%%%%%%%%%%%%%%%%%%%%%%%%%
% \begin{align}
% & 
% \!\!\!
% \mathcal{H}_{D}  =     
%  -t 
%  \sum_{<ij>}^{N_D}\sum_{\sigma}  
%  \left(
%  d^{\dagger}_{i\sigma}d^{\phantom{\dagger}}_{j\sigma}  
% + d^{\dagger}_{j\sigma}d^{\phantom{\dagger}}_{i\sigma}  
% \right) 
% + 
% \sum_{i=1}^{N_D}\sum_{\sigma} \, 
%  \left(\epsilon_d - \frac{H}{2}\,\mbox{sgn}\,\sigma \right)
%  d^{\dagger}_{i\sigma}d^{\phantom{\dagger}}_{i\sigma}  
%  +  U\sum_{i=1}^{N_D} 
%    d^{\dagger}_{i \uparrow}
%    d^{\phantom{\dagger}}_{i \uparrow}
%    d^{\dagger}_{i \downarrow}
%    d^{\phantom{\dagger}}_{i \downarrow} .
% \label{eq:HC^U}
% \\
% &
% \!\!\!
% \mathcal{H}_{\rm mix} =   
% v  \sum_{\sigma}   
%  \left(  
%   d^{\dagger}_{1,\sigma} \psi^{\phantom{\dagger}}_{L \sigma}
%    +   
% d^{\dagger}_{N_D, \sigma} \psi^{\phantom{\dagger}}_{R\sigma}
% + \mathrm{H.c.}
%  \right)   ,
% \label{eq:Hmix}
%  \quad \ 
% \mathcal{H}_{\rm lead}  =  
% \sum_{\nu=L,R} 
%  \sum_{k\sigma} 
% \left(
%   \epsilon_{k}^{\phantom{0}}
% - \frac{H}{2}\,\mbox{sgn}\,\sigma \right)
%          c^{\dagger}_{k \nu \sigma} 
%          c^{\phantom{\dagger}}_{k \nu \sigma}
% .
% \end{align}
%%%%%%%%%%%%%%%%%%%%%%%%%%%%%%%%%%%%%%%%%%%%%%%%%%%%%%%%%%%%%%%%
Here, $t$ $>0$ is the hopping matrix element between the dots,
%along the triangle, 
 $\epsilon_d$ the onsite energy in the dots, and $U$ 
the Coulomb interaction.  
The conduction electrons around the dots, 
 $C_{\nu \sigma}^{\phantom{\dagger}} 
\equiv \sum_k c_{k \nu \sigma}^{\phantom{\dagger}}/\sqrt{N}$, 
can tunnel into the triangle via $v$, 
and it causes the level broadening $\Gamma \equiv \pi  \rho v^2$ 
with $\rho$ the density of states of the leads.
In the limit of $v=0$ and $U=0$,   
the circular motion along the triangle forms 
a single orbital at $E_a \equiv -2t + \epsilon_d$ and 
two degenerate orbitals at $E_b \equiv t + \epsilon_d$. 
This degeneracy brings interesting varieties to the Kondo effect,
occurring in the triangular quantum dot at different values 
of $\epsilon_d$, which can be controlled by the gate voltage. 
% To be specific, we choose the hopping matrix element such that $t>0$ 
% in the following. 

%  that we describe in the following.
%  in the triangular configuration.

%
The low-energy states of the whole system 
including the leads show a local Fermi-liquid behavior,
which is characterized by the two phase shifts,  
$\delta_\mathrm{e}$ and $\delta_\mathrm{o}$, 
for the quasi-particles with the even and odd parities.  
Specifically, at $T=0$, the conductance $g_\mathrm{s}$ and 
the local charge $N_\mathrm{el}$ in the triangle 
can be expressed, respectively, 
in the Landauer and Friedel-sum-rule forms \cite{ONH},
\begin{align}
&\!\!\!
g_\mathrm{s} \,=\, \frac{2e^{2}}{h}\,
\sin^{2} \left(\delta_\mathrm{e} -\delta_\mathrm{o} \right)\;,
\label{eq:gs}
\qquad \quad 
N_\mathrm{tot}
\equiv \,\sum_{i=1}^{N_D}\sum_{\sigma} \,
\langle d_{i\sigma}^{\dagger}d_{i\sigma}^{\phantom{\dagger}}\rangle
\, = \, \frac{2}{\pi}\left(\delta_\mathrm{e}+\delta_\mathrm{o}\right) .
% \label{eq:N_el}
\end{align}
Owing to an inversion symmetry,  
the conductance $g_\mathrm{p}$ for the current flowing 
along the horizontal direction in a four-terminal geometry,
 shown in Fig.\ \ref{fig:system} (b),  
can also be obtained from these two phase shifts 
% $\delta_\mathrm{e}$ and $\delta_\mathrm{o}$ 
defined with respect to the two-channel model $\mathcal{H}$, 
as  $\, g_\mathrm{p} = ({2e^{2}}/{h}) 
     \left(\sin^{2}\delta_\mathrm{e}  + 
     \sin^{2}\delta_\mathrm{o}\right)$.
% \begin{align}
% &\!\!\!\!\!\!\!\!\!\!\!\!\!\!\!\!
% \!\!\!\!\!\!\!\!\!\!\!\!\!\!\!\!
% \!\!\!\!\!\!\!\!\!\!\!\!\!\!\!\!
% \!\!\!\!\!\!\!\!\!\!\!\!\!\!\!\!
% \!\!\!\!\!\!\!\!\!\!\!\!\!\!\!\!
% g_\mathrm{p} \,= \,\frac{2e^{2}}{h} \,
%  \left(\sin^{2}\delta_\mathrm{e} \, + \,
%  \sin^{2}\delta_\mathrm{o}\right)\;.
%       \label{eq:gp}
% \end{align}
%
% can also be calculated 
% from these two phase shifts defined with respect to 
% the series connection Fig.\ \ref{fig:system} (a).
% This is due to the inversion symmetry. 
%
We have deduced the value 
of $\delta_\mathrm{e}$ and $\delta_\mathrm{o}$ 
from the fixed point of NRG \cite{ONH,OH,NO}. Furthermore,
the quantum dots contribution of the free energy $F_D$ 
 may be defined by \cite{KWW},  
\begin{align}
& F_D \, \equiv \, 
F - F_\mathrm{lead} \;,
\qquad \quad 
e^{-F/T} = \mbox{Tr}\, e^{-\mathcal{H}/T}
\;, \qquad 
e^{-F_\mathrm{lead}/T} = \mbox{Tr}\, e^{-\mathcal{H}_\mathrm{lead}/T}
\;. 
% F =
% -T \,\log \left[ \mbox{Tr}\, e^{-\mathcal{H}/T} \right] 
% , \quad 
%F_\mathrm{lead} =
%-T\,\log \left[ \mbox{Tr}\, e^{-\mathcal{H}_\mathrm{lead}/T} 
%\right] 
%
% F_\mathrm{D} \, \equiv \, -T \left[ \,
% \log \left( \mbox{Tr}\, e^{-\mathcal{H}/T} \right) 
% \,-\,
% \log \left( \mbox{Tr}\, e^{-\mathcal{H}_\mathrm{lead}/T} 
% \right) \, 
% \right] 
%
\end{align}
One can also calculate 
the entropy  $\mathcal{S} \equiv -\partial F_D /(\partial T)$ and
spin susceptibility $\chi \equiv -\partial^2 F_D /(\partial H^2)$. 
Here, the magnetic field $H$ is introduced by 
replacing the energy levels $\epsilon_d$ and  $\epsilon_{k}^{\phantom{0}}$, 
respectively, by 
$\epsilon_{d\sigma} \equiv  \epsilon_d
- H\,\mbox{sgn}\,\sigma\,$ and 
  $\,\epsilon_{k\sigma}^{\phantom{0}}\equiv
\epsilon_{k}^{\phantom{0}}
- H\,\mbox{sgn}\,\sigma $.

% \begin{align}
% \epsilon_{d\sigma} \equiv  \epsilon_d 
% - \frac{H}{2}\,\mbox{sgn}\,\sigma \;, 
% \qquad \qquad 
%  \epsilon_{k\sigma}^{\phantom{0}}\,
% \equiv 
%   \epsilon_{k}^{\phantom{0}}
% - \frac{H}{2}\,\mbox{sgn}\,\sigma 
%  \qquad \quad 
% \left\{ \!\!
% \begin{array}{l} 
% \mbox{sgn}\,\uparrow \ \ = +1
%  \\
% \mbox{sgn}\,\downarrow \ \ = -1
% \\
% \end{array}
% \right. 
% .
% \end{align}

% of the orbitals from the leads 

\begin{figure}[t]
 \leavevmode
\begin{minipage}[t]{0.47\linewidth}
 \includegraphics[width=1.0\linewidth]{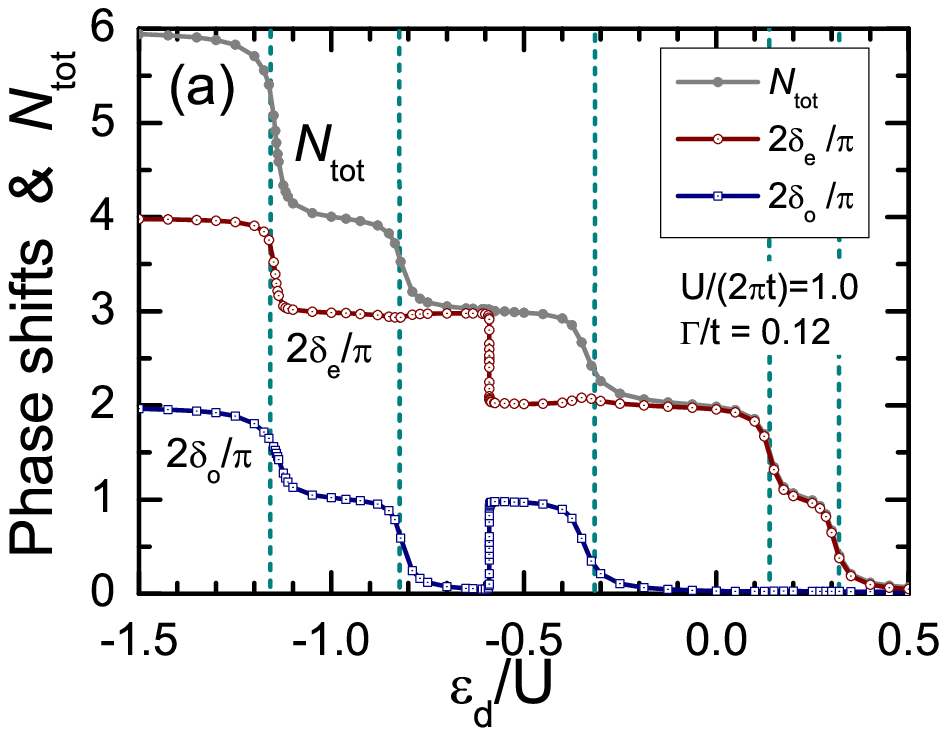}
\end{minipage}
\hspace{0.03\linewidth}
\begin{minipage}[t]{0.48\linewidth}
\includegraphics[width=1.0\linewidth]{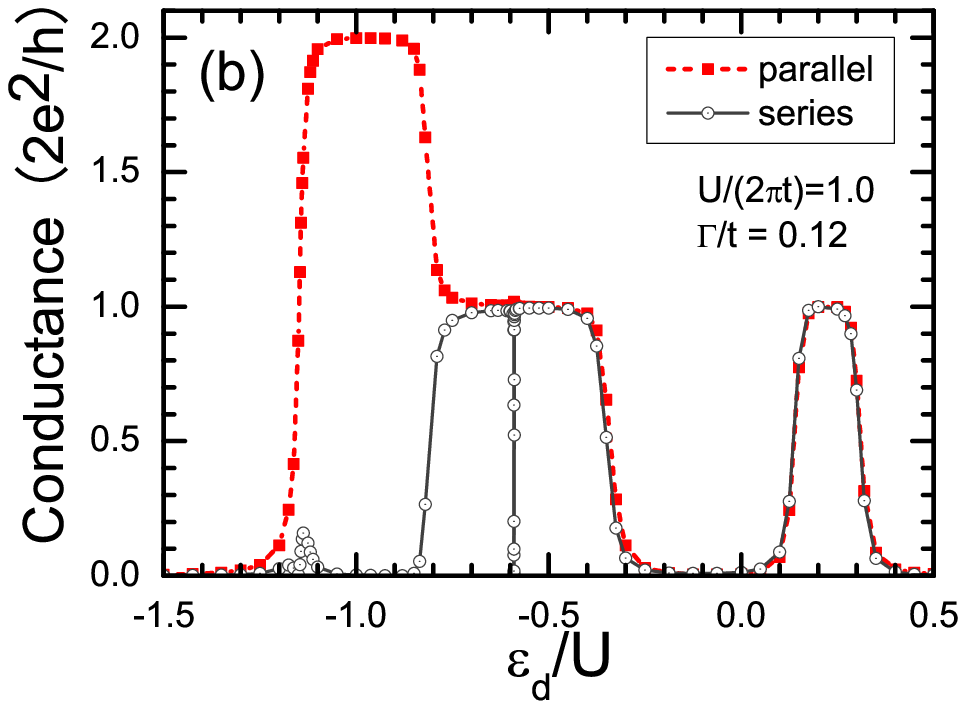}
\end{minipage}
 \caption{
Plots of the NRG results as a function of $\epsilon_d$ for 
 $U/(2\pi t)=1.0$ and $\Gamma/t = 0.12$. 
Left: 
the number of electrons $N_\mathrm{tot}$ in the triangle,
the phase shifts  $2\delta_\mathrm{e}/\pi$ and 
 $2\delta_\mathrm{o}/\pi$. 
The vertical dashed lines 
correspond to the values of $\epsilon_d$, 
at which $N_\mathrm{tot}$ changes discontinuously 
in the isolated limit $\Gamma \to 0$.
Right:  
series ($\odot$, $g_\mathrm{s}$) and parallel 
 ({\protect\small {\color{red} $\fullsquare$}}, $g_\mathrm{p}$) 
conductances. % $\blacksquare$
  
% with $t/D =0.1$ and $\Lambda=6.0$.
 }
 \label{fig:cond_u1}

\end{figure}

\section{NRG results}

In Fig.\ \ref{fig:cond_u1}, 
the results of (a) the phase shifts $\delta_\mathrm{e}$, $\delta_\mathrm{o}$ 
and local charge $N_\mathrm{el}$, 
and (b) the conductances $g_\mathrm{s}$ and $g_\mathrm{p}$,  
are plotted as functions of $\epsilon_d$.
The number of electrons in the triangle $N_\mathrm{el}$ increases 
with decreasing $\epsilon_d$, showing 
the steps for $N_\mathrm{tot} \simeq1,2,3,4,$ and $6$.
It changes directly from $N_\mathrm{el}\simeq 4$ to $N_\mathrm{el}\simeq 6$ 
at $\epsilon_d \simeq -1.15 U$ without 
taking the step for $N_\mathrm{tot} \simeq 5$. 
This is caused by a property of the triangle:   
the ground state becomes degenerate 
for the three charge states $N_\mathrm{tot} =4,\,5$, and $6$ 
at $\epsilon_d \simeq -1.15 U$ in the isolated limit $\Gamma=0$.
The Coulomb repulsion lifts partly 
the degeneracy due to the orbitals with the energy $E_b$.
Specifically at the filling  $N_\mathrm{tot} \simeq 4$,
 a high-spin $S=1$ state becomes the ground state for $U>0$ 
by the Nagaoka mechanism for the ferromagnetism. 
For finite $\Gamma$, the conduction electrons tunneling from the two leads 
screen the $S=1$ moment at low temperatures 
to form a Kondo singlet. 
The series and parallel conductances through the singlet ground state 
show a clear difference in 
this region $-1.15\lesssim \epsilon_d/U \lesssim -0.83$,  
as seen in Fig.\ \ref{fig:cond_u1} (b). 
We see that both of the partial waves contribute to 
the parallel conductance to give the plateau 
of $g_\mathrm{p} \simeq 4e^2/h$, while 
the series conductance is suppressed $g_\mathrm{s} \simeq 0$ 
by the interference. Correspondingly,
the phase shifts take the values $\delta_\mathrm{e} \simeq 3\pi/2$ 
and $\delta_\mathrm{o}\simeq \pi/2$.

%
% it has also been seen 
% in a linear chain of quantum dots \cite{NO,ONH}. 

For small values of the electron occupancies 
 $N_\mathrm{tot} \lesssim 2.0$, seen at  $-0.3 \lesssim \epsilon_d/U$,
the phase shift $\delta_\mathrm{e}$ for the even-parity 
partial wave increases for decreasing $\epsilon_d$, while 
the odd one is almost zero $\delta_\mathrm{o} \simeq 0$. 
%
% We have also calculated directly that 
% the occupation of each of the orbitals 
%
It implies that in this region the electrons occupy mainly 
the orbital of $E_a$ that has an even parity, 
although generally each phase shift does not necessary 
correspond to the occupancy of the eigenstates with the same parity.
Nevertheless we have confirmed, 
by calculating directly the occupation number 
of the each the partial waves  
$N_\mathrm{even}$ and $N_\mathrm{odd}$ which satisfies
 $N_\mathrm{tot}= N_\mathrm{even}+N_\mathrm{odd}$,  
that an approximate relation $N_\mathrm{even} \simeq 2 \delta_\mathrm{e}/\pi$
holds well for each partial wave in the two regions, 
$\epsilon_d/U \lesssim -0.85$ and $-0.6\lesssim \epsilon_d/U$.
However, such a separation of the Friedel sum rule does not 
hold for the intermediate region $-0.85 \lesssim \epsilon_d/U \lesssim -0.6$: 
while each phase shift takes a constant value 
 $2\delta_\mathrm{e}/\pi \simeq 3.0$ and $2\delta_\mathrm{o}/\pi \simeq 0.0$, 
the occupation calculated directly has a different constant  
$N_\mathrm{even} \simeq 2.5$ and $N_\mathrm{odd} \simeq 0.5$. 
These values suggest that, 
after two electrons occupy the orbital of the energy $E_a$, 
the third electron occupies 
the degenerate orbitals of the energy $E_b$ in equal weights. 
Although the coupling with the two leads breaks 
the full triangular symmetry of the point group,
% the effect of the Coulmb interaction is to
it seems as if the charge distribution recovers effectively 
the degeneracy at $\epsilon_d \simeq -0.6 U$. 
% considered to recover the degeneracy 
%
During the sudden change of the phase shifts at $\epsilon_d \simeq -0.6U$, 
the sum $\delta_\mathrm{e} +\delta_\mathrm{o} \simeq 3\pi/2$ 
and the parallel conductance $g_\mathrm{p} \simeq 2e^2/h$ change 
their values very little. 
On the other hand, the phase 
difference $\delta_\mathrm{e} -\delta_\mathrm{o}$, 
varies rapidly from $\pi/2$ to $3\pi/2$.
This rapid and continuous change of  $\delta_\mathrm{e} -\delta_\mathrm{o}$  
causes the narrow dip observed 
for the series conductance $g_\mathrm{s}$ 
at $\epsilon_d \simeq -0.6U$ 
 in the middle of the conductance plateau.
The dip is also related to the SU(4) symmetry.
In contrast, the conductance at $N_\mathrm{tot} \simeq 1$ 
is caused by the usual SU(2) Kondo effect due to $S=1/2$ moment.
%

% \bigskip

\begin{figure}[t]
 \leavevmode
\begin{minipage}[t]{0.47\linewidth}
\includegraphics[width=1.0\linewidth]{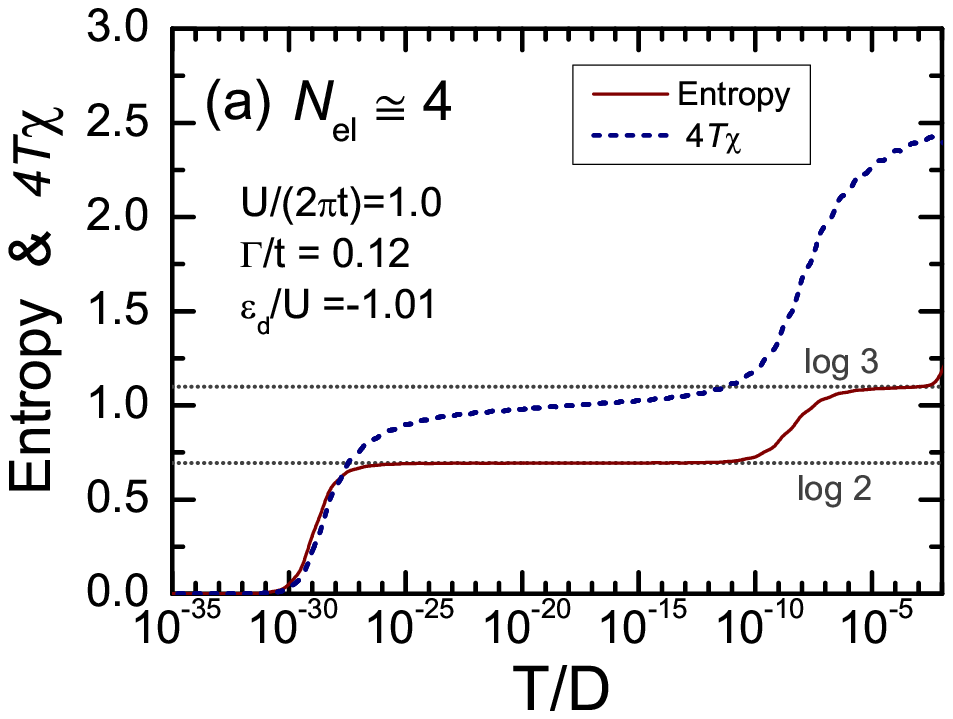}
\end{minipage}
\hspace{0.03\linewidth}
\begin{minipage}[t]{0.48\linewidth}
 \includegraphics[width=1.0\linewidth]{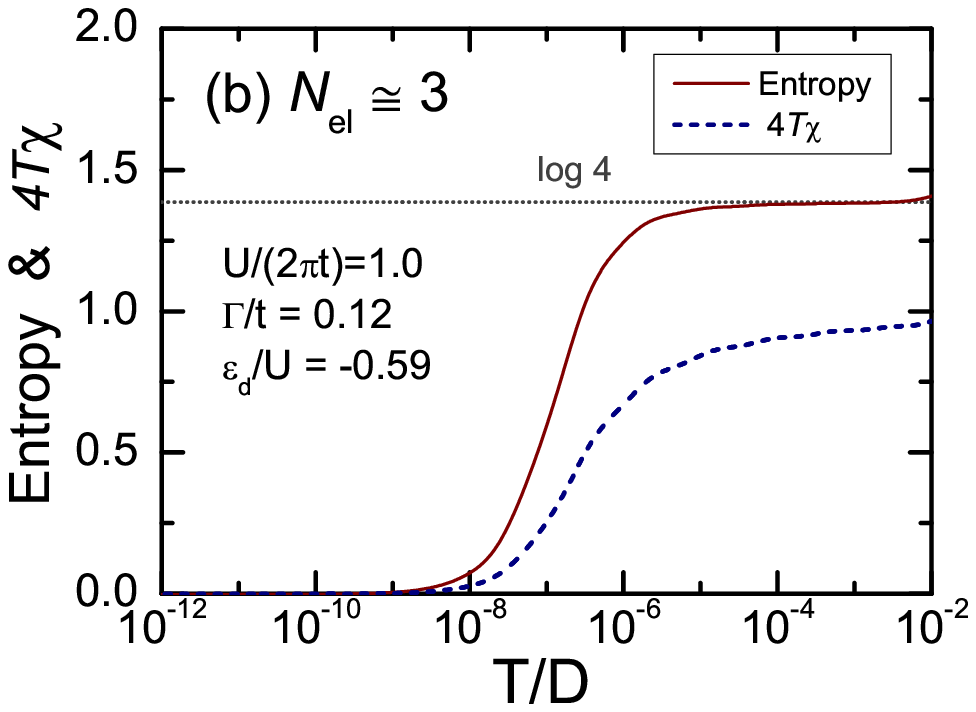}
\end{minipage}
 \caption{
Plots of the triple-dot contribution of entropy  
and $4 T \chi$ as 
a function of $\log (T/D)$ for $U/(2\pi t)=1.0$ and $\Gamma/t = 0.12$.
Here, $\chi$ is the spin susceptibility, and $D$ is the half-width of 
the conduction band.
Left (a): results at $\epsilon_d = -1.01U$ ($N_\mathrm{tot}\simeq 4$), 
where the triangle has the $S=1$ moment.
Right (b): results at $\epsilon_d = -0.59U$ ($N_\mathrm{tot}\simeq 3$), 
where $g_s$ shows the sharp dip.
% $\blacklozenge$ $\blacksquare$ 
% $t/D =0.1$ and $\Lambda=6.0$.
 }
 \label{fig:S_Tchi}

\end{figure}

In Fig.\ \ref{fig:S_Tchi} (a) the temperature dependence of 
the entropy $\mathcal{S}$ and $4 \chi T$, which are obtained 
at $\epsilon_d = -1.01U$ in the middle of the charge step 
for $N_\mathrm{tot} \simeq 4$, 
is plotted as a function of $\log(T/D)$,
where $D$ is the half-width of the conduction bands. 
The results clearly show that the screening of the spin $S=1$ 
local moment due to the Nagaoka mechanism occurs 
via two separate stages at $T/D \simeq 10^{-10}$ and $T/D \simeq 10^{-30}$. 
We see that the entropy takes the value $\mathcal{S}\simeq\log 3$
at high temperatures  for $T/D \gtrsim 10^{-7}$, and it  
 means that the $S=1$ moment is free.
Then, 
at intermediate temperature  $10^{-28} \lesssim T/D \lesssim 10^{-10}$,  
the entropy becomes $\mathcal{S}\simeq\log 2$. 
It shows that  
the half of the moment is screened by the conduction electrons from 
one of the channel degrees, and the local moment is still in 
an under-screened state. 
The full screening is completed 
finally at very low temperatures $T/D \lesssim 10^{-30}$,
as we can seen also in the behavior of $\chi$.

Similarly, the entropy and the spin susceptibility 
for $\epsilon_d = -0.59U$,  
 just at the dip of $g_\mathrm{s}$ for $N_\mathrm{tot} \simeq 3$,
are plotted in Fig.\ \ref{fig:S_Tchi} (b). 
At the high temperatures $T/D \gtrsim 10^{-6}$, 
the entropy takes a constant value $\mathcal{S} \simeq \log 4$,
which implies that there exist four degenerate local states 
at the triangle. 
The screening completes at $T/D \simeq 10^{-8}$, 
and it may be regarded as the SU(4) Kondo effect.
We see that the spin susceptibility  
for $T/D \gtrsim 10^{-4}$ shows 
the Curie behavior $\chi \simeq S(S+1)/(3T)$ with 
the coefficient corresponding to $S=1/2$.
Therefore the degenerate states have the spin $S=1/2$ degrees 
of freedom, and thus the remaining degrees of freedoms 
of the four fold degeneracy should have 
an orbital origin in the triangle.

\section{Summary}
We have studied the ground state properties of 
a triangular triple quantum dot connected 
to two noninteracting leads, and have found  
that the system shows the SU(4) and 
the $S=1$ two-stage Kondo effects, 
as well as the usual SU(2) one,
depending on the gate voltage  $\epsilon_d$.

%\smallskip
\ack
We would like to thank J.~Bauer and V.~Meden for valuable discussions.
This work was supported 
by JSPS Grant-in-Aid for Scientific Research (C).
Numerical computation was partly carried out 
in Yukawa Institute Computer Facility.

\end{document}